# Traffic Optimization for TCP-based Massive Multiplayer Online Games

Jose Saldana, Luis Sequeira, Julián Fernández-Navajas, José Ruiz-Mas
Communication Technologies Group (GTC) – Aragon Inst. of Engineering Research (I3A)
Dpt. IEC. Ada Byron Building. CPS Univ. Zaragoza
50018 Zaragoza, Spain
{jsaldana, lsequeirav, navajas, jruiz}@unizar.es

*Abstract*— **This paper studies the use of a traffic optimization technique named TCM (Tunneling, Compressing and Multiplexing) to reduce the bandwidth of MMORPGs (Massively Multiplayer Online Role-Playing Games), which employ TCP to provide a soft real-time service. In order to optimize the traffic and to improve bandwidth efficiency, TCM can be applied when the packets of a number of players share the same link, which occurs in some scenarios, as e.g. the traffic between proxies and servers of game-supporting infrastructures. First, TCP/IP headers are compressed using standard algorithms that avoid sending repeated fields; next, a number of packets are blended into a bigger one and finally, they are sent using a tunnel. The expected compressed header size has been obtained using traffic traces of a real game. Next, simulations using a traffic model of a popular MMORPG have been performed in order to estimate the expected bandwidth savings and the reduction in packets per second. The obtained bandwidth saving is about 60 percent. Packets per second are also significantly reduced. In addition, the added delays are shown to be small enough so as not to impair players' experienced quality.**

## I. INTRODUCTION

Massively Multiplayer Online Role-Playing Games (usually known as MMORPGs) are becoming very popular in the last years. They create a virtual world which is simultaneously shared by thousands of players, each of them controlling an avatar. The players have to accomplish some missions, for which they need to collect different objects. Initially, they have some powers and weapons, which can be improved as they reach higher levels.

These games are acquiring more and more popularity, and the most well-known of them (*World of Warcraft*, developed by Blizzard Ent.) claims to have more than 10 million users worldwide [1]. It reached a maximum number of 12 million users, although the raising of other titles which are free to play is reducing its share. Nevertheless, it can still be considered the MMORPG par excellence.

Taking into account the high number of users, the problem of supporting the service after the release of a title is not trivial. The success of a game is not totally predictable and, in fact, some games, as *Diablo II*, had serious supporting problems in the first months after their release [2]. The problem becomes worse since game players have been reported to be very difficult to satisfy, i.e., if a server does not match their requirements, they would leave and never return [3].

As a result, game providers have to design and deploy suitable supporting infrastructures, with enough processing capacity and bandwidth. A hierarchical structure of game servers, some of them acting as proxies, has been proposed in the literature [4]. The gamers follow certain connection patterns during the day, and this means that there are some critical moments when the number of simultaneous players becomes very high [1]. Some studies have shown that game servers cannot easily share their capacity with other services (e.g. web) [3], as they present similar daily periodic workload peaks. Consequently, techniques providing bandwidth and workload savings are interesting so as to avoid the need of over provisioning the resources. In the same study, the existence of a limit in the packets per second (pps) that a router can manage was highlighted, and it was recommended to consider this pps limit it in addition to bandwidth limit.

Although MMORPGs are between the most popular online games, there are other genres that are also played by millions of users. As an example, First Person Shooters (FPSs) are also a consolidated group of games with some characteristics in common: the virtual scenario is shared by a few tens of players, who use fire weapons to kill the enemies or accomplish a mission. The weapon can be improved as the player earns money, depending on their fighting skills.

The main differences between these two genres can be summarized [5], [6], [7] as follows:

- Session duration in MMORPGs is longer than in FPSs. Nevertheless, FPS gamers usually play a number of rounds during the same session.
- The number of players sharing a virtual scenario in an FPS is of a few tens, whereas thousands of players can simultaneously share the virtual world of an MMORPG.
  The real-time and interactivity requirements of FPSs are higher, and the good aim of the player is of primary importance. In contrast, MMORPGs are not based on the good aim, since the players first use the mouse to select the objective, and then they choose the weapon or the curse to use against the enemy.
- FPSs use UDP protocol, while most MMORPGs use TCP.

This work has been partially financed by CPUFLIPI Project (MICINN TIN2010-17298), MBACToIP Project, of Aragon I+D Agency and Ibercaja Obra Social, and NDCIPI-QQoE Project of Catedra Telefonica, Univ. of Zaragoza.

The last difference is the most important for the current study: the use of UDP in FPSs means that there is no retransmission when a packet is lost. This fact has some implications: e.g. a shot can be lost, so players usually use machine-gun bursts so as to kill the enemies. Some games implement packet loss concealment algorithms in order to hide network impairments to the players [8]. The latency is of primary importance in these games, as remarked in [9]. In fact, network latency can make a player miss a hit which was initially good[1].

On the other hand, MMORPGs normally use TCP, which is reliable and avoids the loss of any information related to players' actions. When a packet is lost, the protocol asks for a retransmission. But the actions, and especially the fights, do have some interactivity: the player can select different weapons and curses while the fight is on, and their ability to do this quickly has a significant impact on the final result. So we can conclude, as reported in [5], that these games can be considered as a new class of service: (soft) real-time and interactive using TCP.

In order to avoid additional delays, MMORPGs usually set to 1 the "push" flag of TCP header [5], thus forcing the packet to be sent as soon as possible. As a result, they tend to generate small TCP packets. This is a very interesting scenario where header compression can be applied: we have long-term flows, with many header fields that are the same for every packet.

This paper presents the use of a Tunneling, header Compression and Multiplexing technique (TCM), in the context of MMORPG traffic. These techniques were first developed for RTP flows [10], where significant savings can be achieved [11]. In [12] the technique was adapted to UDP flows of FPS games, and significant savings were also obtained. The reduction in terms of packets per second was also high. In the current study we try to find whether the technique is also suitable to compress TCP flows of MMORPGs, taking into account the special issues that appear when multiplexing is applied to a reliable protocol.

Regarding the scenarios of application, the technique can be first used between the servers that support the game (Fig. 1 a): a proxy may receive the traffic of all the users of a zone (e.g. a town or a district), and forward it to the central server, so the aggregated traffic between the proxy and the server can be compressed and multiplexed, while adding small delays. This may provide some flexibility to the supporting infrastructure: when bandwidth gets scarce, traffic is multiplexed; and when the number of users diminishes, the traffic is sent in its native form, thus avoiding additional delays.

Other scenarios, as an Internet Café (Fig. 1b), where a number of users simultaneously play the same game, can also be suitable for this technique. However, in this case the number of users may be significantly smaller than the one for the server-to-server scenario. In the tests we will have to determine the number of players for which the use of the technique is interesting, taking into account that subjective quality must not be harmed.

---

[1] Some expert players compensate the effect of network latency aiming the weapon some meters ahead their enemy, when he is moving, performing a sort of "correction of the point of aim".

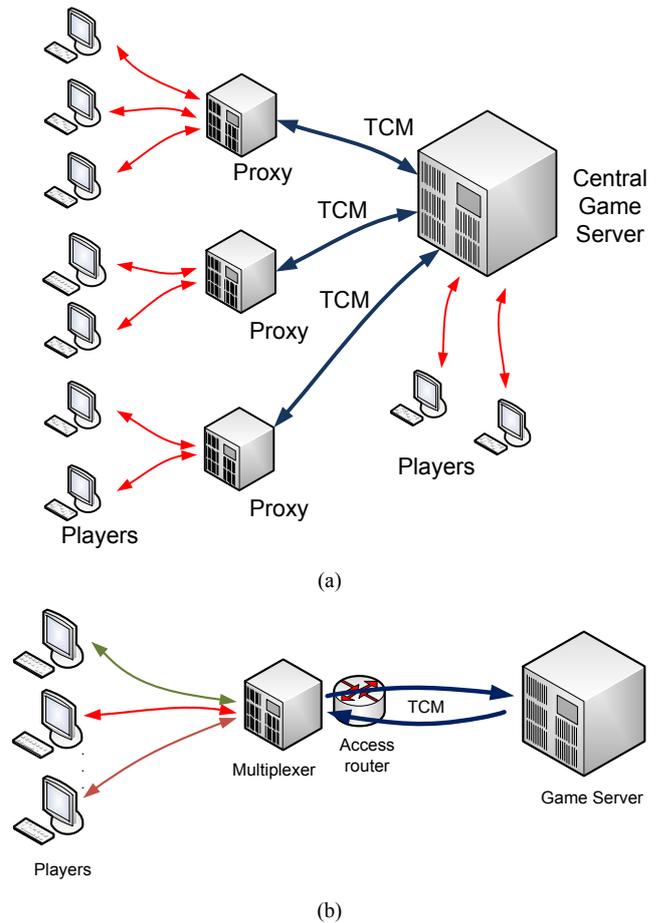

Figure 1. Scenarios where TCM can be applied: a) traffic between servers; b) Internet café.

The structure of the paper is as follows: related works are discussed in section II. Next, the use of the compressing and multiplexing technique for MMORPG traffic is studied. Section IV presents the test methodology and the final results. The paper ends with the Conclusions.

II. RELATED WORKS

Many topics have to be considered in this section. First, the use of proxies for game-supporting infrastructures was proposed in [2] and [13], taking into account the stringent requirements of this concrete service. In [14] the feasibility of a peer-to-peer support for MMORPGs was studied, and one of the conclusions was that message aggregation can reduce network latency.

Regarding header compression methods, a number of them have been defined and standardized. They mainly avoid the sending of the header fields that are the same for every packet of a flow, and they also use delta compression for reducing the number of bits of changing fields. This requires the use of a *context,* which stores the value of non-changing header fields, e.g. IP addresses and ports. Logically, context synchronization between the origin and the destination is of primary importance.

The first method for compressing TCP/IP headers was proposed by Van Jacobson [15]. Later, IPHC [16] also included the possibility of compressing IPv6 and UDP headers. At the same time, cRTP was defined [17], being also capable of compressing IP/UDP/RTP headers. Some years later, ECRTP [18] presented some improvements with respect to cRTP in links with high delay, packet loss and packet reordering. The last compressing algorithm presented was ROHC [19], which prevents the desynchronization of the context, especially in wireless scenarios.

Multiplexing methods were first designed for RTP flows, due to the existence of scenarios where a number of real-time flows may share the same path. The IETF defined TCRTP as RFC 4170 [10], in order to compress headers, also using PPPMux so as to include a number of native packets into a multiplexed one. Finally, an L2TPv3 tunnel was included in order to permit the end-to-end sending of packets. This method was adapted for its use when the traffic is not RTP/UDP but only UDP [12], showing its ability to obtain significant bandwidth savings, which could be above 30% for IPv4, and 50% when IPv6 was considered. In the current article, we study the feasibility of this method when applied to (soft) real-time TCP flows.

A number of traffic models have been developed for FPS games [20]. The traffic of MMORPGs has also been analyzed in [5], [7]. These studies, based on traffic traces of a game, deploy a mathematical model, which is compared to the original one by means of suitable analytical tools, as Q-Q plot. The developed models allow the generation of synthetic traffic, which can be useful for further research, thus avoiding the need of playing the game while performing network measurements.

Finally, it must be said that subjective quality models have also been developed for online games. They were first developed for VoIP [21], but they have also been adapted for different games. The final result of the model is a MOS (Mean Opinion Score) value, which ranges from 1 (bad) to 5 (excellent). The threshold value of acceptable quality is usually considered to be about 3.5. The problem is that each game presents a different behavior with respect to each concrete network parameter, since different techniques are used by developers for the concealment of network impairments [6]: for example, in [8] it was reported that, while the players of *Quake IV* are surprisingly not aware of packet loss up to 35 %, Microsoft's *Halo* stops working when packet loss is 4%. As a consequence, each game has to be particularly studied by means of subjective surveys. In [22] a subjective quality model for *World of Warcraft* was presented, based on delay and jitter.

## III. ANALYSIS OF TCM FOR MMORPG TRAFFIC

In this section, the behavior of TCM is firstly explained. Next, the header compression protocol is more deeply studied. The formula of the expected bandwidth saving when using TCM for this traffic is presented in the last subsection.

### A. Summary of TCM and TCP-related issues

Traffic compressing and multiplexing can be applied to TCP/IP flows, as shown in Fig. 2. It consists of three steps: first, a header compression algorithm is applied to the headers of the packets. Next, compressed packets are multiplexed using PPPMux; and finally, the bundle is sent to the destination using PPP and L2TPv3 tunneling, which allows end-to-end delivery.

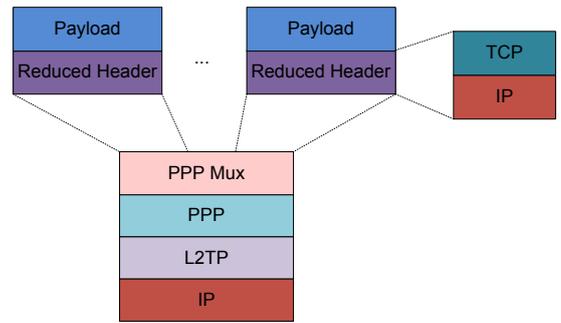

Figure 2. TCM protocol stack, for TCP/IP header compressing.

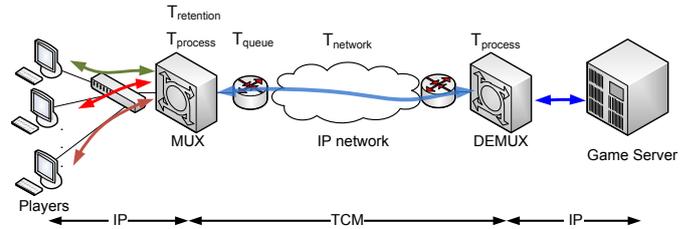

Figure 3. Scheme of the elements of the system and their associated delays.

The technique becomes interesting when a number of flows share the same path. This permits the sharing of the common header between all the packets, taking into account that the header size will be slightly increased due to the use of the tunnel. As shown in Fig. 3, the tunnel is only established from the multiplexer to the demultiplexer, where the packets are rebuilt exactly as they were generated by the application. This means that multiplexing is transparent to the game and to the server, so it can be independently applied for client-to-server and/or server-to-client traffic.

Different policies, which were compared in [23], can be used so as to select the packets to multiplex in each bundle. In the present study, we have used a policy that defines a period, and blends all the packets arrived to the multiplexer, sending them at the end of the interval (Fig. 4), thus establishing an upper bound for the added delay. As a result, packets from different flows share the same multiplexed bundle. In order to avoid packets bigger than MTU (Maximum Transmission Unit), a size threshold is also defined, and a multiplexed packet is sent whenever that threshold is reached, even if the period has not finished.

Finally, we will do some considerations about the specific problems that appear when using this scheme with TCP flows, taking into account that the multiplexing scheme was developed for real-time flows using UDP packets. If we look at Fig. 2, we can see that the multiplexed bundle is not carried over TCP, so TCP payloads are being transmitted over an unreliable protocol. If a packet loss occurs, then all the multiplexed packets will be lost. As a result, the retransmission mechanisms of TCP will act for each of the flows. Therefore, multiplexing is transparent for the communication ends, and it can be seen as an additional delay.

Regarding packet loss probability, the number of packets sent is reduced by a factor of $E[k]$, i.e. the average number of multiplexed packets. However, the loss of a multiplexed packet is equivalent to the loss of *E*[*k*] native ones. As a result, the packet loss probability will remain the same.

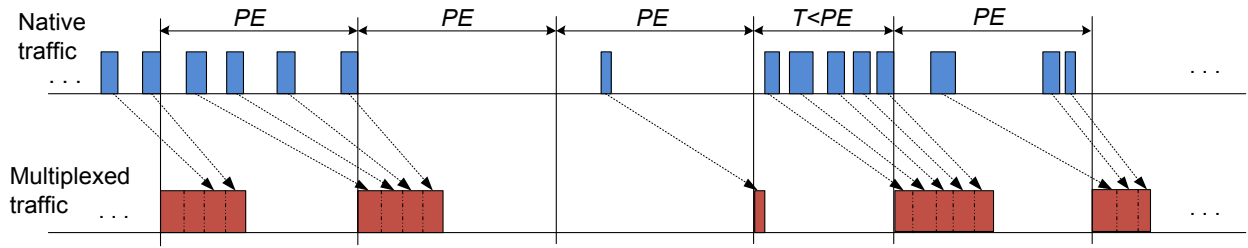

Figure 4. Multiplexing policy using a period *PE*

This can make us question if it could be more adequate to add a new TCP layer between the IP and L2TP layers of the tunnel. This second TCP layer would retransmit multiplexed lost packets. But we have decided not to use this option on behalf of simplicity: while a multiplexed bundle is being retransmitted, the retransmission mechanisms of some of the native TCP flows may also act, asking for new packets which would be subsequently multiplexed.

### B. Header Compression Algorithm

Different header compression protocols can be used in TCM. We need a protocol capable of compressing TCP/IP headers, so we may select IPHC or ROHC. Although they use similar compression methods, the latter has been designed to perform well even in links with high RTT and packet loss, as it happens in wireless environments. It sacrifices some amount of compression so as to improve context synchronization guarantees [24]. As the scenarios considered in the present work are wired networks with a very low packet loss rate, IPHC is considered more adequate for our proposal.

In order to obtain the expected size of the header, we will briefly summarize the IPHC algorithm, which was adapted from [15], and jointly compresses TCP and IP headers.

The protocol sends two different header types:

- FULL_HEADER: it establishes or refreshes the context of a packet stream, represented by a *context identifier* (CID). It presents the same size of the original, but it includes the CID value in the second byte of the *total length* field of the IP header. The length of the packet is inferred from lower layer protocols.
- COMPRESSED_TCP: in the rest of the cases, a compressed packet is sent. Its scheme can be seen in Fig. 5. The first byte includes the identifier of the context (CID), and the second one is a mask that indicates which fields are present in the header, e.g. if the bit *S* is set to 1, this means that the field *Δsequence (S)* is present. There is an exception: the bit *P* is a copy of the one of the original header. This is the *push* bit, and it indicates that this packet has to be sent immediately.

The fields that are the same for every packet of the flow are denoted as *DEF* fields, and they are only included in full headers. *Random* fields are the ones for which delta compression is not suitable, since they change randomly. They have to be sent in each packet, and they are included after *TCP checksum*, in the same order as they appear in the original header.

Ref. [15] defined a mechanism for including full fields instead of delta ones when necessary: if 8 bits are not enough to express the change in the field (i.e. a change bigger than 256), then an extra byte of zeros is included, and next, the full field. So a decision has to be made, depending on the behavior of a field: if the number of times it significantly changes is big, then it will be better to include it as a *random* field, thus avoiding the additional byte of zeros. The probabilities of having each header size have to be computed in order to obtain the expected value of the compressed TCP/IP header for each concrete application.

### C. Expected Savings

In order to obtain a formula for the expected bandwidth savings, we can calculate the expected size of the packets arrived in a period, and also the expected size of the compressed packet. For simplicity, we will not consider the effect of the MTU limit. Thus, we will use the next variables:

- *NH*: The native header size: 40 bytes for TCP/IPv4.
- *CH*: The size of the common header, which will be 25 bytes if IPv4 is used: 20 for IP header, 4 for L2TP and 1 for PPP.
- *MH*: PPPMux header (2 bytes).

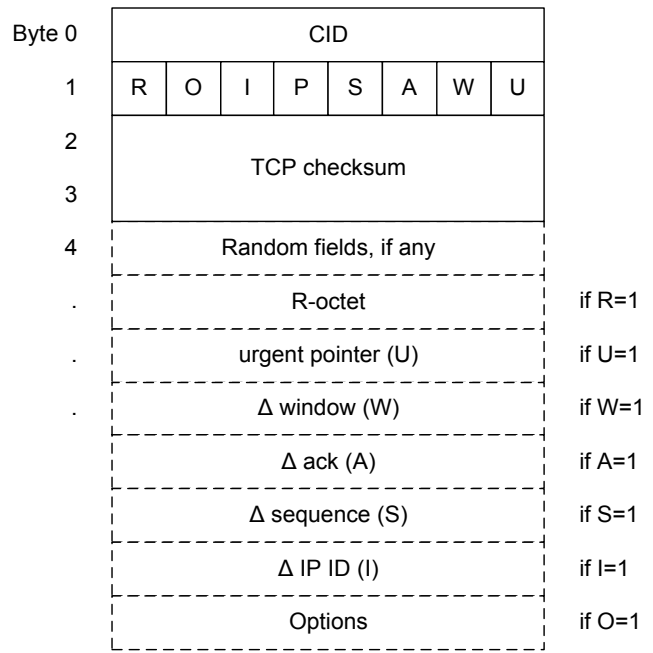

Figure 5. Header of a COMPRESSED_TCP packet

- $E[P]$: The expected value of the TCP payload, which depends on the application. It must be taken into account that ACK packets without payload can also be compressed, so they will also be considered in the calculation of the expected value of the payload, with a value of 0. This will make the calculations depend on the TCP implementation and parameters of the game server and the machine of the player.

- $E[k]$: The average number of native packets included into a multiplexed one.

- $E[RH]$: The expected value of the reduced header. We will calculate it for the studied application in the next section.

The expected size in bytes of native and multiplexed packets arrived in a period will respectively be:

$$bytes_{native} = E[k] (NH + E[P]) \quad (1)$$

$$bytes_{mux} = CH + E[k] (MH + E[RH] + E[P]) \quad (2)$$

As a consequence, the bandwidth saving *(BWS)* can be obtained as:

$$BWS = 1 - \frac{bytes_{mux} / period}{bytes_{native} / period} =$$

$$= 1 - \frac{CH}{E[k](NH + E[P])} - \frac{MH + E[RH] + E[P]}{NH + E[P]} \quad (3)$$

The second term is caused by the sharing of the common header, and it is reduced as the number of multiplexed packets grows. The third term represents the relationship between compressed and native headers of each packet. It will have a fixed value for each title, so an asymptote for the maximum bandwidth saving will appear.

## IV. TEST METHODOLOGY AND RESULTS

In the tests, we have mainly used *World of Warcraft*, because of the next reasons: first, it is the most popular MMORPG; second, its traffic behavior is typical of this genre; and third, it has been largely studied in the literature [5], [14], and a MOS model has even been developed for it [22].

Header compression and multiplexing can be seen as independent processes, i.e. a flow can be compressed without using multiplexing, and also many native flows can be multiplexed without considering header compression. In this section we will first study the behavior of the header compression algorithm, using real traces of the application under test. Next, the simulation method used for obtaining the multiplexed traffic traces will be summarized. Finally, the results will be presented.

### A. Behavior of the Compression Algorithm

In this subsection, a statistical distribution for the reduced header size will be obtained, allowing us to calculate $E[RH]$. Traffic traces of the game, obtained from real parties performed in our laboratory, have been used in order to obtain the model. We have used a wired connection with a very low packet loss rate, from a Windows 7 64 bits client, to Blizzard's servers[2], and captured 4,000 packets on each direction. The behavior of the header fields that are not the same for every packet has been observed and characterized so as to obtain the probability of having each one of the possible header sizes:

- *R-octet*: It will always be set to 0 for the studied application.
- *Urgent pointer (U):* It is always set to 0, so it will never be present.
- *Δ window (W):* Taking into account that the window size can be increased or reduced, its variation has to be between -127 and 128 in order to be suitable to be expressed in 8 bits. This depends on the TCP stack of the machine of the player. When 8 bits do not suffice, three bytes have to be sent.
- *Δ ack (A):* Its original size is 32 bits, so it may be sent using 5 bytes (including the one of zeros), but in most cases it can be compressed to 8 bits or even avoided.
- *Δ sequence (S):* Its behavior is similar to the one of the *A* field.
- *Δ IPID (I):* It increases by one with respect to the previous packet, so we can use a single byte to represent it.

Some fields present a different behavior for client-to-server and server-to-client traffic. The behavior has been summarized in Table I, which includes the obtained percentages for *W, A* and *S* fields. We can observe that defining these fields as *random* is not interesting: as an example, for *S* field, we have two options: if we define it as *random*, we will always need 4 bytes. If not, we will use a single byte when possible, or five bytes in other case. If we look at the average values obtained, it is better to do it the second way, as the percentages of having no change or a single-byte one are high.

As a result, the header size for client-to-server packets will range from 4 to 14 bytes, whereas server-to-client ones vary from 4 to 11 bytes. As the maximum size of the compressed header is 14 bytes, we could compress headers even if only one packet has arrived, as the addition of the common header plus the compressed one will have an upper bound of 39 bytes. The obtained percentages for each header size are presented in Fig. 6. These values will be used to calculate the size of the compressed headers in subsequent simulations. The expected size of the reduced header is 8.72 bytes for client-to-server packets and 7.37 bytes for server-to-client ones. Taking into account that the original TCP/IP header uses 40 bytes, it can be observed that the saving is significant.

TABLE I.   BEHAVIOR OF THE HEADER FIELDS

| field | no change | 8 bits | full | average number of bits |
|---|---|---|---|---|
| *client to server* | | | | |
| W | 17.58 % | 62.24 % | 20.18 % | 1.224 |
| A | 17.41 % | 50.91 % | 31.68 % | 2.093 |
| S | 59.47 % | 40.53 % | 0 % | 0.405 |
| *server to client* | | | | |
| W | 100 % | 0 % | 0 % | 0 |
| A | 65.55 % | 34.45 % | 0 % | 0.341 |
| S | 20.56 % | 48.38 % | 31.06 % | 2.034 |

---

[2] Although some private servers (e.g. *TrinityCore*) have been developed for *World or Warcraft*, they are unofficial and based on reverse engineering.

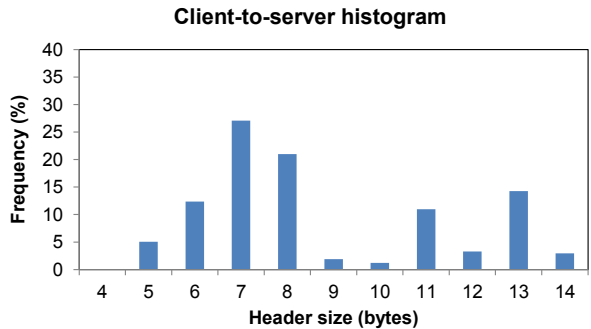

(a)

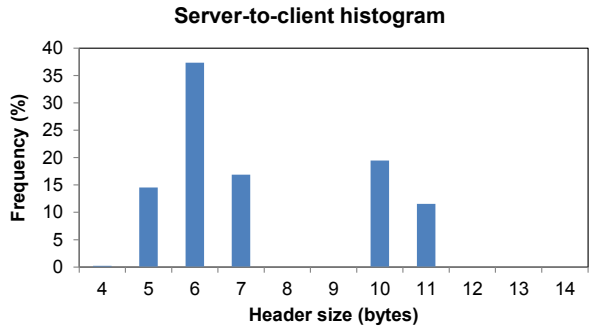

(b)

Figure 6. Header size distribution of the compressed traffic.

TABLE II. *BWR* ASYMPTOTE VALUES FOR DIFFERENT GAMES

| client to server | | | |
|---|---|---|---|
| game | *E[P]* | pps | max. saving |
| WoW | 8.74 | 9.51 | 60.07 % |
| ShenZhou | 25 | 8 | 45.1 % |
| RoM | 33 | 4.17 | 40.1 % |
| server to client | | | |
| game | *E[P]* | pps | max. saving |
| WoW | 314 | 6.05 | 8.65 % |
| ShenZhou | 114 | 8 | 19.88 % |
| RoM | 99 | 5.17 | 22 % |

Once the expected values for the compressed header sizes have been obtained, we are now able to present some numerical results (Table II) of the bandwidth saving asymptote, as obtained in (3). In order to get a more general idea of the savings which can be obtained for different games, we will calculate the value or the asymptote not only for *World of Warcraft* (WoW), but also for two more MMORPGs: *ShenZhou Online* [7], by UserJoy Technology; and *Runes of Magic* (RoM), by Runewalker Entertainment[3]. Assuming that we are using the same TCP implementation, we can use the calculated value of *E[RH]* for the three games.

It can be observed that the bandwidth saving is significant for client-to-server traffics, whereas it is lower for server-to-client ones. Concretely, in the case of client-to-server traffic of *World of Warcraft,* the maximum saving is roughly 60 percent of the bandwidth. On the other hand, the server-to-client saving for this game is very small, because of the big size of the packets. The values for the other games are roughly 20 percent.

---
[3] The values for this game are based on preliminary measurements which have been deployed by our group. Although a complete traffic model has not already been presented, the average packet size can be easily obtained.

### B. Obtaining of Traffic Traces

Once the header compression algorithm has been statistically characterized, Matlab simulations have been performed so as to obtain packet sizes and packet departure times of native and multiplexed flows. The process can be divided into three stages (Fig. 7), which we will next explain.

*1) Traffic Generation:* A model for the traffic of *World of Warcraft* was developed in [5], using real traces obtained in real Internet accesses, and also using the traffic generated during one week in a mobile core network. Inter-packet time was modeled by a joint distribution of three uniformly distributed variables. The size of the APDU (Application Protocol Data Unit) was modelled using a Weibull distribution for the downlik, and three possible sizes at the uplink. For the obtaining of this model, the authors first removed the ACK packets having no payload, which were 56% of the uplink packets and 28% of the downlink ones. It must be taken into account that the game also sends ACKs in packets with payload.

We have used Matlab to generate traffic traces for different numbers of players (the number of packets generated is the product of 5,000 and the number of players), using this model in three steps:

- The APDU and inter-packet times are generated.
- If the APDU is bigger than the MTU, it is divided into a number of packets, which are sent in a burst.
- TCP ACK packets without payload are added, using the rates reported in the model, and the correspondent inter-packet time distribution.

*2) Compression of the Flows:* IPHC algorithm, which has been explained above, is applied to each traffic flow, using the statistics obtained in the previous subsection. Although packets from different players are included into the same multiplexed bundle, IPHC is separately applied to each flow. The reason for this is that there is a context that is used to compress each flow.

*3) Multiplexing and Tunneling:* A policy based on a period is used, as shown in Fig. 4. All the packets that arrive during a period are multiplexed together, despite the flow they belong. A threshold of 1,350 bytes is set so as to avoid multiplexed packets bigger than the MTU.

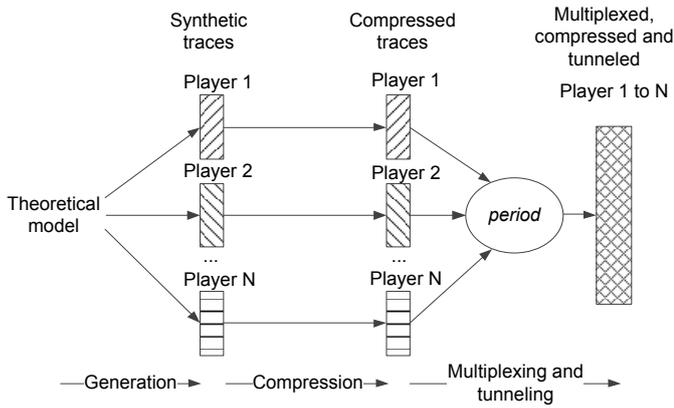

Figure 7. Stages of traffic generation

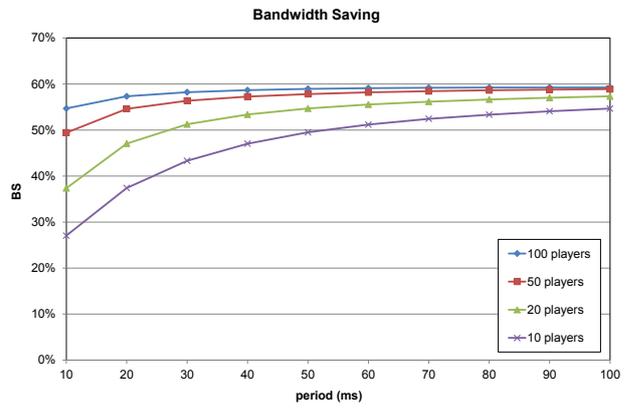

Figure 9. Bandwidth saving for client-to-server traffic of *World of Warcraft*.

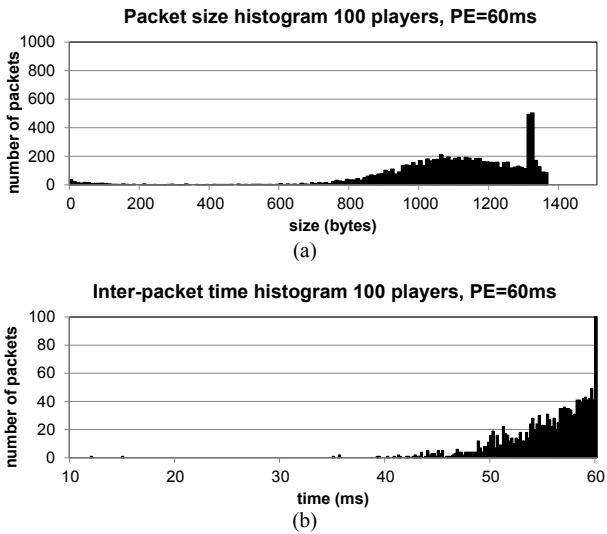

Figure 8. Multiplexed traffic: a) packet size; b) inter-packet time histogram. A peak of 7,462 packets in 60ms has been cut for clarity.

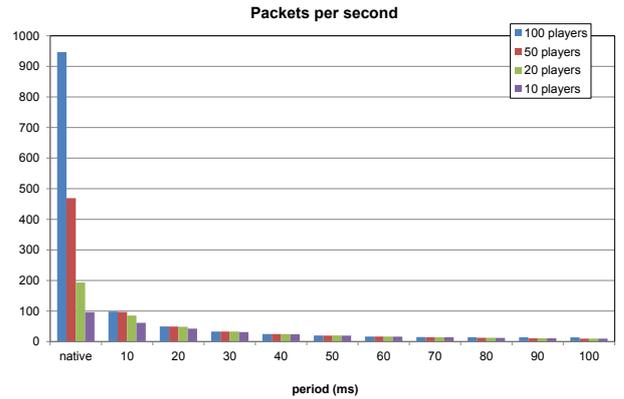

Figure 10. Packets per second for client-to-server traffic of *World of Warcraft*.

Fig. 8 shows packet size (a) and inter-packet time (b) histograms for the multiplexed traffic of 100 players using a period of 60 ms. It can be observed that the method increases packet size, presenting a peak above the threshold of 1,350 bytes. Regarding inter-packet time, there is a peak of 7,462 packets in 60 ms, which has been cut for clarity. This means that many periods last 60 ms.

*C. Results*

In this subsection we present the results, mainly in terms of bandwidth saving and packets per second. We have seen that client-to-server savings are the highest ones, so we mainly study this traffic. Fig. 9 shows the bandwidth saving for different numbers of players, with a period ranging from 10 to 100 ms. We can observe that the curves present an asymptote around 60 %, as expected. If the number of players is small, a period of 50 ms has to be used so as to obtain bandwidth savings above 50 %. Nevertheless, a saving of 25 % can be easily achieved, using a tiny period of 10 ms. On the other hand, when a big number of players' traffic is multiplexed, bandwidth savings of about 50 % can be obtained even for very small values of the period. Fig. 10 presents the packets per second, which can be reduced from 900 to 10, which is the inverse of the multiplexing period.

If we compare the results with the ones obtained using FPS traffic [12], we can see that bandwidth savings can be obtained more easily with MMORPGs, due to the higher compression level of the headers, and also to the presence of TCP ACK packets without payload, in which header compression is directly translated into size reduction. Regarding the number of players, it is possible to have higher numbers, as in these games the scenario is shared by thousands of them. But we have to study the network impairments, which are the counterpart of the multiplexing method, i.e., the added delays, and the jitter, since a packet that arrives at the beginning of the period will be delayed more than a packet arriving at the end.

In order to make sure of not harming players' experience, we have used a subjective quality estimator [22], to build a graph (Fig. 11) using network delays of 20, 40 and 100 ms, with a stdev of 10 ms. These can be considered typical Internet delays for different intra-region or inter-region scenarios [25].

The impairments of multiplexing are added to the delay and jitter of the network. It can be observed that for small network delays, MOS values above 3.5 can be easily achieved. But if the delay is 100 ms, then the period has to be maintained under 50 ms, in order to grant a good user's experience. We see that the players of MMORPGs can tolerate longer delays than FPS's ones, as the interactivity of the game is smaller.

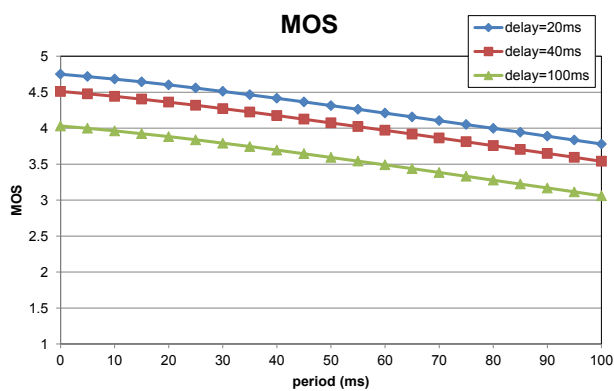

Figure 11. MOS calculated with different network delays and a stdev of 10 ms

## V. CONCLUSIONS

A multiplexing and header compression method has been applied to the traffic of MMORPGs, in scenarios where the flows of a number of players share the same path. The method has been analyzed, showing that there is an asymptote that establishes an upper bound for the bandwidth saving.

The behavior of the header compression algorithm has been studied using real traces of the concrete application, obtaining a statistical characterization of its performance. With these results, simulations using a statistical model of a commercial game have been conducted in order to obtain the average bandwidth saving as a function of the number of players and the multiplexing period. The obtained savings are significant, and can be about 60 percent. An important reduction in the amount of packets per second can also be observed. Finally, by the use of a subjective quality estimator, it has been shown that the system is able to maintain an acceptable experienced quality. Further research could be performed, using the traffic and quality models of other games, in order to build a comparative study between different titles. Different network scenarios including delay and packet loss would also have to be considered, in order to test the effect of these network impairments on the subjective quality of the gamers.